# Tunable orbital angular momentum in high-harmonic generation

D. Gauthier[1,*], P. Rebernik Ribič[1], G. Adhikary[2], A. Camper[3], C. Chappuis[4],
R. Cucini[1], L. F. DiMauro[3], G. Dovillaire[5], F. Frassetto[6], R. Géneaux[4], P. Miotti[6],
L. Poletto[6], B. Ressel[2], C. Spezzani[7], M. Stupar[2], T. Ruchon[4], and G. De Ninno[2,1,*]

1. Elettra-Sincrotrone Trieste, Area Science Park, 34149 Trieste, Italy
2. Laboratory of Quantum Optics, University of Nova Gorica, 5001 Nova Gorica, Slovenia
3. Department of Physics, The Ohio State University, Columbus, Ohio 43210, USA
4. LIDYL, CEA, CNRS, Université Paris-Saclay, CEA Saclay, 91191 Gif-sur-Yvette, France
5. Imagine Optic, 91400 Orsay, France
6. Institute of Photonics and Nanotechnologies, CNR-IFN, 35131 Padova, Italy
7. Laboratoire de Physique des Solides, Université Paris-Sud, CNRS-UMR 8502, 91405 Orsay, France
*. Corresponding Authors: david.gauthier@elettra.eu, giovanni.de.ninno@ung.si

(Dated: August 30, 2016)

**Optical vortices are currently one of the most intensively studied topics in optics. These light beams, which carry orbital angular momentum (OAM), have been successfully utilized in the visible and infrared in a wide variety of applications. Moving to shorter wavelengths may open up completely new research directions in the areas of optical physics and material characterization. Here, we report on the generation of extreme-ultraviolet optical vortices with femtosecond duration carrying a controllable amount of OAM. From a basic physics viewpoint, our results help to resolve key questions such as the conservation of angular momentum in highly-nonlinear light-matter interactions, and the disentanglement and independent control of the intrinsic and extrinsic components of the photon's angular momentum at short-wavelengths. The methods developed here will allow testing some of the recently proposed concepts such as OAM-induced dichroism, magnetic switching in organic molecules, and violation of dipolar selection rules in atoms.**

Angular momentum is a fundamental property of the photon, together with energy and linear momentum. In paraxial conditions, angular momentum may be split into an intrinsic part, the spin angular momentum, and an extrinsic part called orbital angular momentum (OAM). Macroscopically, the OAM of light manifests itself in the spatial properties of a beam and in particular in the shape of its wavefront. The most common light beams carrying OAM are the Laguerre-Gaussian modes, solutions of the wave equation in the paraxial regime. They show an azimuthal phase dependence $\exp(-i\ell\varphi)$ [1,2], where $\varphi$ is the azimuthal coordinate in the transverse plane and $\ell$, called topological charge, is indexing the mode. This phase shape creates a beam with a helical wavefront, a phase singularity in its center and a donut-shaped intensity profile. A topological charge $\ell$ results in an OAM per photon equal to $\ell\hbar$ [1].

These modes, also called optical vortices, have been used in the visible range in many diverse applications. Along with its other degrees of freedom, a photon pair can exhibit OAM entanglement [3,4], which can be exploited in quantum cryptography. The donut-shaped intensity profile of an optical vortex allows bleaching the outer annular part of a sample and, applying stimulated emission depletion microscopy, allows to bypass the diffraction limit in optical microscopy [5,6]. Laguerre-Gaussian modes are also the tool of choice to manipulate particles [7,8] or detect spinning objects [9]. Recently, increasing attention has been devoted to theoretical studies of fundamental interactions involving OAM beams in the extreme-ultraviolet (XUV) spectral range. For example, it was shown that OAM might be transferred to electronic degrees of freedom [10,11]. In Ref. [12], it was also theoretically demonstrated that an XUV vortex can induce charge current loops in fullerenes with an associated orbital magnetic moment, which can be controlled by tuning the topological charge of the incident beam. These findings, if confirmed experimentally, could lead to new applications in magnetic switching using structured light.

These and other theoretical studies have sparked new experimental developments in the field of intense femtosecond XUV pulse generation. Schemes have been proposed for generating optical vortices using free-electron lasers [13,14], which are still to be demonstrated. As a table-top and largely tunable alternative to large scale instruments, high-harmonic generation (HHG) is based on the frequency upconversion of a high intensity femtosecond Vis-IR laser into the XUV range through a highly-nonlinear process [15]. Recently, OAM has been studied in the context of HHG in gases. The first experimental work reported on the difficulty of generating optical vortices due to the non-conservation of the OAM during propagation in the gas jet [16]. Subsequent experiments, however, demonstrated the generation of optical vortices carrying a topological charge that is a multiple of the harmonic order [17,18], in agreement with the expected conservation rule for a single driving beam [19]. Nevertheless, with such experimental schemes, the topological charge could not be tuned independently of the harmonic order. In addition, only high values of the topological charge could be obtained, and no XUV beam with unitary topological charge was demonstrated. These two restrictions will severely limit the applicability of the above schemes in most of the recently proposed experiments.

In this Article, we report on an alternative scheme to produce optical vortices carrying an arbitrary topological charge for any harmonic order using HHG. Our setup is based on a two-color wave-mixing arrangement that combines a Gaussian beam with a frequency doubled Laguerre-Gaussian beam in a gas target, as proposed in ref. [17]. We use a Hartmann sensor to measure the helical wavefront of the generated optical vortices. We also study the intensity scaling of the different modes regarding the usual HHG parameters. Our study provides the first experimental verification of the conservation rule for OAM in HHG using two driving beams. Moreover, the use of two-color driving beams allows the efficient and robust generation of optical vortices carrying a low topological charge (from $\ell$ = 1 to $\ell$ = 4). This relatively simple setup makes HHG the first light source of femtosecond XUV pulses carrying a controllable amount of OAM.

The experiment is sketched in Fig. 1 and was performed on the CITIUS light source, described in more details elsewhere [20]. A Ti:sapphire laser system (5 kHz repetition rate) provides pulses with a duration of $\simeq$ 50 fs and energy of $\simeq$ 2mJ at a central wavelength of 805 nm. The pulses are sent through a 450 µm thick type-I BBO crystal, giving the fundamental (ω) and second harmonic (2ω) beams. The two spectral components are then spatially separated in a Mach-Zehnder-like interferometer equipped with dichroic mirrors. The interferometer is used to manipulate individually the two beams before being focused by two independent lenses into a 1 mm long argon gas cell. In one arm, the 2ω beam is converted from a Gaussian to a Laguerre-Gaussian mode with $\ell_{2\omega} = 1$ using a spiral phase plate. No OAM is imparted on the ω beam ($\ell_\omega = 0$). The crossing angle between the two beams in the gas cell can be adjusted by translating the last mirror of each arm. All modes generated at the frequency of one harmonic order of the fundamental frequency ω are isolated out from the other harmonic orders by a monochromator consisting of a grating placed between two toroidal mirrors, which provides an image of the HHG modes at the position of the slits located further downstream. The grating is used in the so-called conical geometry (diffraction in the vertical direction) to limit the spatial frequency dispersion (spatial chirp) of the imaged harmonic beam and the pulse front tilt [20]. Finally, the intensity distribution and the wavefront are measured in the far field using an XUV CCD or a Hartmann wavefront sensor (developed by Imagine Optic, SOLEIL, and Laboratoire d'Optique Appliquée [21]).

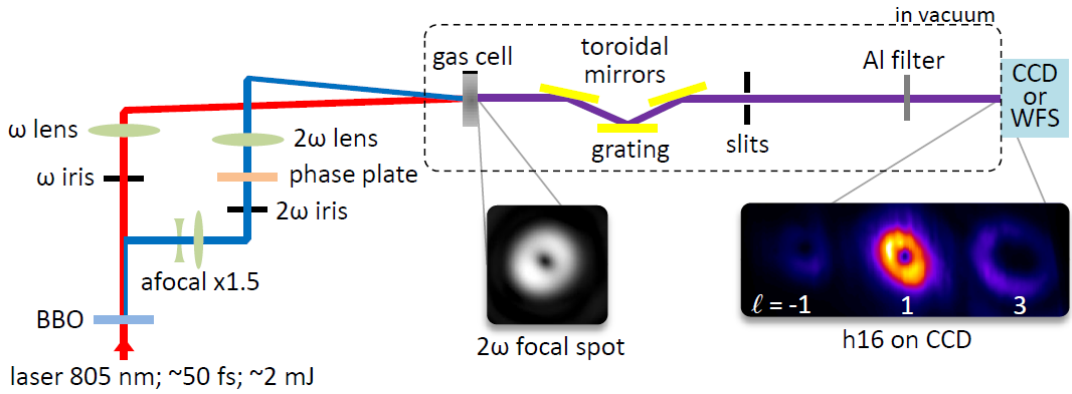

*Figure 1: Scheme of the experimental setup. The collimated laser beam (8 mm in 1/e²) goes through a BBO crystal before entering a dichroic Mach-Zehnder interferometer in which the depleted 805 nm spectral part is independently manipulated from the frequency doubled part. The spiral phase plate converts the Gaussian beam into a Laguerre-Gaussian mode with $\ell_{2\omega} = 1$ in the second harmonic (2ω) arm of the interferometer, while the fundamental (ω) beam remains Gaussian. The mirrors and the delay line used to synchronize the two beams are not drawn, nor the half wave plate used to get identical linear polarizations. Both beams are focused with lenses of 75 cm focal length into the generation gas cell ($\simeq 1$ mm length) filled with a few mbar of argon. To optimize the focusing geometry, both lenses are placed on translation stages, irises are positioned in both arms and a beam expander (x1.5) is used in the 2ω arm. The XUV emission is sent into a monochromator (made of a grating between two toroidal mirrors and slits) before being imaged on an XUV CCD camera or a Hartmann wavefront sensor (WFS) placed in the far-field, $\simeq 1$m downstream. (Insets) Typical intensity distributions at focus of the 2ω beam (left) and HHG vortices (right, labeled by their topological charge $\ell$) generated at the $16^{th}$ order imaged on the CCD.*

Figure 2 shows the far-field intensity profile of different harmonic orders obtained using a ω-2ω crossing angle of 13 mrad. The spatial axis origin is taken along the propagation axis of the ω beam. For each high-harmonic order, we observe the angular distribution (in the horizontal direction) of spatially separated "modes". The emission angle of each mode is determined by the non-collinear phase-matching condition, in agreement with the conservation of energy, linear momentum and parity, previously established for HHG in the two-color configuration [22,23]. It is convenient to use the photon picture to describe the possible pairs ($n_1$, $n_2$) that contribute to the emission of a given high-harmonic order q (with q = $n_1$ + 2 $n_2$), where $n_1$ and $n_2$ refer respectively to the number of photons absorbed from the ω and 2ω beams. In particular, parity requires that only the absorption of an odd total number of photons n = $n_1$ + $n_2$ can lead to emission. Consequently, in Fig. 2 we observe modes generated from the pairs (19, 0), (15, 2) and (11, 4), contributing to the emission of photons with an energy corresponding to the $19^{th}$ harmonic order (h19). For h18, the pairs experimentally observed are (20, -1), (16, 1) and (12, 3). Note that only the fundamental beam is contributing to the generation of the (q, 0) pair, while generation with the fundamental alone is not allowed for even harmonics. Positive and negative numbers in pairs are related, respectively, to sum- and difference-frequency generation.

The intensity patterns in Fig. 2 are consistent with the conservation of OAM that will be confirmed with the phasefront measurement. This result is in agreement with the theoretical model developed in Ref. [17], and demonstrates the transfer of OAM from the generating beams to high harmonics. For a two-color wave mixing setup, we can generalize the conservation rule for OAM as:

$$\ell = n_1 \ell_\omega + n_2 \ell_{2\omega},$$

where $\ell$, $\ell_\omega$ and $\ell_{2\omega}$ are the topological charge carried, respectively, by the high-harmonic, the ω and the 2ω beams. In the present experiment we focus on the case $\ell_\omega = 0$ and $\ell_{2\omega} = 1$, which gives simply $\ell = n_2$. Therefore, the topological charge carried by each mode is equal to the number of 2ω photons absorbed in the process. Consequently, except for pairs with $n_2$ = 0, all modes display a ring-like intensity profile with zero intensity in the center, characteristic of optical vortices. Moreover, for each harmonic order, the size of the rings increases with the number of absorbed 2ω photons, consistent with the increase of the topological charge as demonstrated in Ref. [18].

The results in Fig. 2 are representative of the main experimental finding for our generation conditions. For the low harmonic orders, in some cases, the modes in the far field are composed of inner and outer parts. Note that this feature exists for ℓ=0 with or without the presence of the 2ω driving beam. Such spatial profiles can be attributed to the contributions of two different quantum paths to the high-harmonic dipole of emission in combination with propagation effects [24-26]. The second noticeable feature is the spatial distortion of the mode profiles in the far field, which is due to aberrations in the XUV transport optics (monochromator) and to aberrations of the generating beams. Despite major efforts to reduce such aberrations, a residual astigmatism was still present on the generating beams (visible in the focus of the 2ω beam, Fig. 1). Aberrations in the generating beams drive (generally stronger) aberrations in the HHG emission, and constitute one cause of the non-conservation of OAM. Indeed, astigmatism in the generating beams can convert Laguerre-Gaussian modes into Hermite-Gaussian ones, due to the astigatic mode conversion [27,1]. A residual effect related to this is visible in Fig.2 at h13 for the ℓ=2 mode. When the modes are not fully spatially separated, the far-field intensity pattern also contains slight interference features.

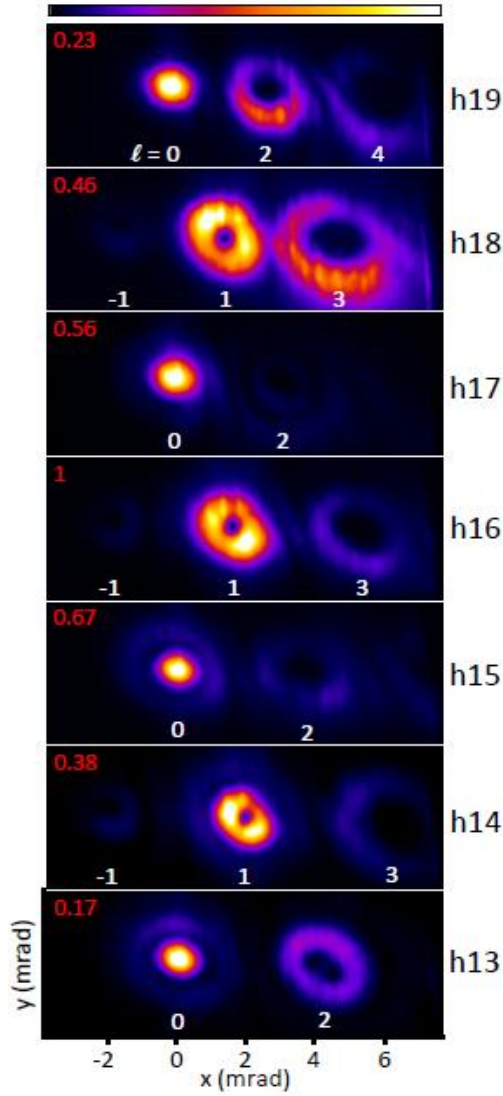

*Figure 2: Far-field intensity patterns obtained using the non-collinear wave-mixing scheme described in Fig. 1. Single shot images for different harmonic orders (hq) were taken with the CCD using identical generating conditions. Each mode is labeled by its topological charge ℓ that also corresponds to the number of absorbed 2ω photons (n2). For better visualization, images for each harmonic order are shown with the full color scale (color bar on top). The integrated signal (top left corner) is normalized to h16. The number of photons incident on the CCD (after the metallic filter) for h16 was around $7 \times 10^6$. Note that the angular acceptance of the monochromator (about 10 mrad) limited the number of modes that could be detected at once to three.*

To confirm that the observed rings are actual (quasi) Laguerre-Gaussian modes, for odd and even harmonics, we performed wavefront measurements using a Hartmann wavefront sensor (WFS). Some representative results are displayed in Fig. 3. On the top-left panel, we clearly show a spiraling wavevector around a singularity, as detected by the WFS. The magnitude of the wavevector increases when it gets closer to the center of the beam. These two observations are characteristic of an optical vortex. With our convention, we have a left-handed vortex for ℓ>0. The phase maps displayed in the other panels in Fig. 3 were obtained from the integration of the wavevector distribution. A reference wavefront (taken without the spiral phase plate) was used to correct for the phase aberrations due to the monochromator optics. We measured very smooth phases spiraling around the beam propagation axis going from 0 to (almost) ℓ×2π radians. Note that the small discrepancies in the maximum variation of the phase across the map are attributed to a relatively low sampling of the wavefront, except for the case ℓ=4. This measurement is a direct and unambiguous characterization of the OAM carried by the vortex beams, including its sign. It also demonstrates the possibility of directly measuring helical phasefronts in the XUV spectral range, which will be instrumental in the development of new generation light sources such as free-electron lasers carrying OAM.

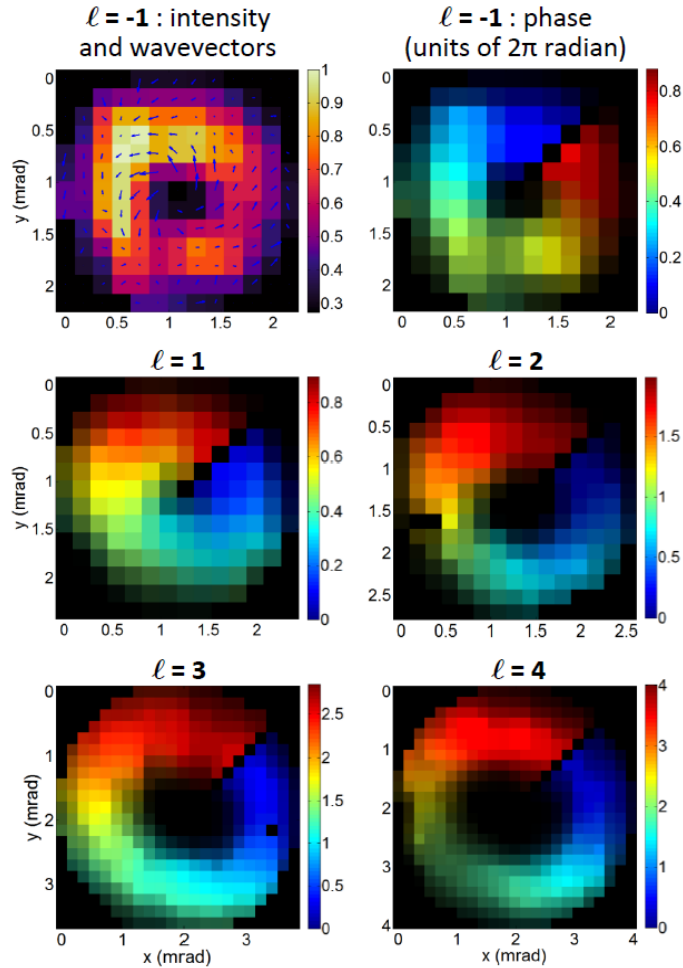

*Figure 3: Intensity and wavefront measurements for several modes carrying various topological charges. The top-left panel shows the intensity (in arbitrary units) and the local wavevector (arrows) distribution for the ℓ=-1 mode of the 16$^{th}$ high-harmonic order emission (h16). The other panels show the measured wavefronts for ℓ=-1 and ℓ=1 for h16, ℓ=2 fot h19, ℓ=3 for h18 and ℓ=4 for h19. The intensity is represented by the brightness and the phase (in units of 2π radian) by false colors. Note that the different panels have different color scales. The generation conditions were optimized for each mode, and a larger non-collinear angle than the one in Fig.2 was used to guarantee a better spatial separation between the modes.*

For a given high-harmonic order, the signal is not equally distributed between the modes. This is a known effect in highly non-linear wave mixing [23,28], due to both the microscopic response of the medium and the macroscopic effects (propagation and phase matching) in HHG. This property can be exploited to optimize and favor the emission of a specific vortex by modifying the generation conditions. Figure 4 shows the evolution of the generated signal among 3 modes of the 16$^{th}$ harmonic order when varying the iris aperture/transmitted energy in each arm of the interferometer and the gas cell pressure. In Fig. 4 (a), the iris apertures modify the intensity as well as the size of both beams at focus. These parameters impact the individual atomic response in the gas jet (extension of the generating area, amplitude and phase of the dipoles), the phase matching conditions, and the propagation and reshaping of the fundamental beams in the medium [29,30]. The intensity ratio of the second harmonic to the fundamental beam (2ω/ω) was varied from a few percent up to 50 percent, spanning both perturbative and non-perturbative regimes. This relative intensity between the two colors is the relevant parameter to explain the evolution of the signal within each mode. Fig. 4 (b) shows how the gas pressure impacts the macroscopic effects during HHG and influences the global efficiency of generation [31,30], as well as the non-collinear phase matching [28]. The latter can explain the variation of the relative signal distribution between the 3 modes. Finally, note that an optimized gas pressure allows generating an ℓ=3 vortex with a flux of $8.5 \times 10^{10}$ photons/s. This value is competitive with regard to some synchrotron beamlines in the same spectral range.

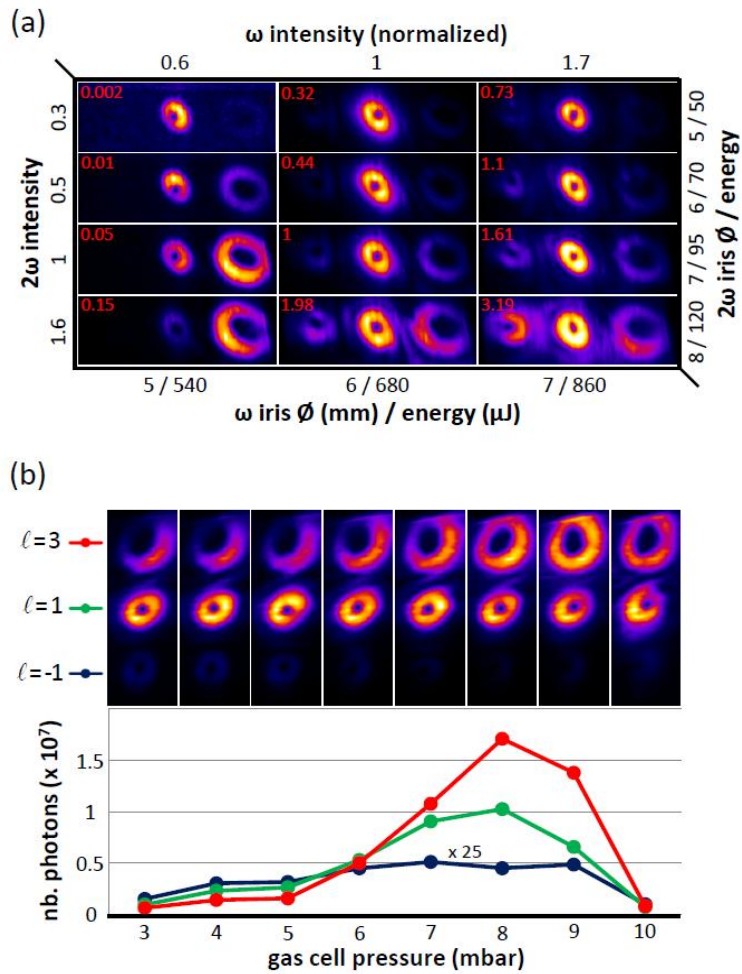

*Figure 4: Evolution of the modes as a function of the intensity of the generating beams at focus (a) and gas pressure (b) for the 16th harmonic order. (a) The intensity of the generating beams at focus is normalized to an estimated intensity ratio $I_{2\omega}/I_{\omega}$ of about 18%, with $I_{\omega} \sim 1.7 \times 10^{14}$ W/cm². The intensity is calculated from the measured iris aperture and transmitted energy. The signal is normalized to $1.7 \times 10^{7}$ photons per shot on the CCD (top left corner of each image). (b) The curves display the yield for each individual mode, together with the corresponding image.*

In conclusion, we demonstrated the generation of XUV vortex beams with tunable orbital angular momentum. Our findings enhance the capabilities of spatial shaping of HHG light compared to schemes using a single generating beam [32]. The method we propose can be naturally extended to any combination of generating beams with various spatial properties. As an example, similar wave-mixing schemes have been applied to the generation of circularly polarized high harmonics using two-color Gaussian beams with various polarization states [33,34]. Combining one of these schemes with the method presented in this work will allow generation of femtosecond HHG pulses with independent control over orbital and spin angular momenta. This will in turn pave the way towards new fundamental experiments in the field of light-matter interactions, such as the study of the coupling between the extrinsic and intrinsic angular momenta of light.

*Acknowledgments:*

This work was partially supported by the Bilateral collaboration Slovenia-Commissariat à l'énergie atomique et aux énergies alternatives (CEA, France) 2016-2018, sponsored by the Slovenian Research Agency (ARRS) and by CEA .This work was supported by the French Agence Nationale de la Recherche (ANR) through XSTASE project (ANR-14-CE32-0010). AC and LFD acknowledges support of the US Department of Energy, Office of Science, Office of Basic Energy Sciences under contract DE-FG02-04ER15614.

*References:*

[1] L. Allen, M. W. Beijersbergen, R. J. C. Spreeuw, and J. P. Woerdman, Phys. Rev. A 45, 8185 (1992)

[2] M. Padgett, J. Courtial, and L. Allen, Phys. Today 57(5), 35 (2004)

[3] A. Mair, A. Vaziri, G. Weihs and A. Zeilinger, Entanglement of the orbital angular momentum states of photons, Nature 412, 313-316 (2001)

[4] G. Molina-Terriza, J. P. Torres and L. Torner, Twisted photons, Nature Physics 3, 305-310 (2007)

[5] A. Jesacher, S. Furhapter, S. Bernet, and M. Ritsch-Marte, Phys. Rev. Lett. 94, 233902 (2005)

[6] Stefan W. Hell and Jan Wichmann, Breaking the diffraction resolution limit by stimulated emission: stimulated-emission-depletion fluorescence microscopy, Optics Letters 19, pp. 780-782 (1994)

[7] T. Kuga, Y. Torii, N. Shiokawa, and T. Hirano, Novel optical trap of atoms with a doughnut beam, Phys. Rev. Lett., vol. 78, pp. 47134716, (1997)

[8] H. He, M. Friese, N. R. Heckenberg, and H. Rubinsztein-Dunlop, Direct observation of transfer of angular momentum to absorptive particles from a laser beam with a phase singularity, Phys. Rev. Lett., vol. 75, pp. 826-829, (1995)

[9] M. P. J. Lavery, F. C. Speirits, S. M. Barnett, and M. J. Padgett, Science 341, 537 (2013)

[10] A. Picón, J. Mompart, J. R. Vázquez de Aldana, L. Plaja, G. F. Calvo, and L. Roso, Photoionization with orbital angular momentum beams, Optics Express 18, 3660-3671 (2010)

[11] O. Matula, A. G. Hayrapetyan, V. G. Serbo, A. Surzhykov and S. Fritzsche, Atomic ionization of hydrogen-like ions by twisted photons: angular distribution of emitted electrons, J. Phys. B: At. Mol. Opt. Phys. 46, 205002 (2013)

[12] J. Waetzel, Y. Pavlyukh, A. Schaeffer, J. Berakdar, Optical vortex driven charge current loop and optomagnetism in fullerenes, Carbon 99, 439e443 (2016)

[13] E. Hemsing, A. Knyazik, M. Dunning, D. Xiang, A. Marinelli, C. Hast and J. B. Rosenzweig, Coherent optical vortices from relativistic electron beams, Nature Physics 9, 549–553 (2013)

[14] P. Rebernik Ribič, D. Gauthier and G. De Ninno, Generation of Coherent Extreme-Ultraviolet Radiation Carrying Orbital Angular Momentum, PRL 112, 203602 (2014)

[15] Thomas Brabec and Ferenc Krausz, Intense few-cycle laser fields: Frontiers of nonlinear optics, Reviews of Modern Physics, 72 (2000)

[16] M. Zürch, C. Kern, P. Hansinger, A. Dreischuh and Ch. Spielmann, Strong-field physics with singular light beams, Nature Physics 8, 743–746 (2012)

[17] G. Gariepy, J. Leach, K. T. Kim, T. J. Hammond, E. Frumker, R. W. Boyd, and P. B. Corkum, Creating High-Harmonic Beams with Controlled Orbital Angular Momentum, Phys. Rev. Lett. 113, 153901 (2014)

[18] R. Géneaux, A. Camper, T. Auguste, O. Gobert, J. Caillat, R. Taïeb, T. Ruchon, Attosecond light and electronic vortices, arXiv:1509.07396 (2015)

[19] Carlos Hernández-García, Antonio Picón, Julio San Román, and Luis Plaja, Attosecond Extreme Ultraviolet Vortices from High-Order Harmonic Generation, Phys. Rev. Lett. 111, 083602 (2013)

[20] C. Grazioli, C. Callegari, A. Ciavardini, M. Coreno, F. Frassetto, D. Gauthier, D. Golob, R. Ivanov, A. Kivimäki, B. Mahieu, B. Bučar, M. Merhar, P. Miotti, L. Poletto, E. Polo, B. Ressel, C. Spezzani and G. De Ninno, CITIUS: An infrared-extreme ultraviolet light source for fundamental and applied ultrafast science, Rev. Sci. Instrum. 85, 023104 (2014)

[21] Pascal Mercère, Philippe Zeitoun, Mourad Idir, Sébastien Le Pape, Denis Douillet, Xavier Levecq, Guillaume Dovillaire, Samuel Bucourt, Kenneth A. Goldberg, Patrick P. Naulleau, and Senajith Rekawa, Hartmann wave-front measurement at 13.4 nm with lambda/120 accuracy, Optics Letters 28, 1534-1536 (2003)

[22] M. D. Perry and J. K. Crane, Phys. Rev. A 48, R4051 (1993)

[23] J. B. Bertrand, H. J. Wörner, H.-C. Bandulet, É. Bisson, M. Spanner, J.-C. Kieffer, D. M. Villeneuve, and P. B. Corkum, Ultrahigh-Order Wave Mixing in Noncollinear High Harmonic Generation, Phys. Rev. Lett. 106, 023001 (2011)

[24] Pascal Salières, Anne L'Huillier, and Maciej Lewenstein, Coherence Control of High-Order Harmonics, Phys. Rev. Lett. 74, 3776 (1995)

[25] Philippe Balcou, Pascal Salières, Anne L'Huillier, and Maciej Lewenstein, Generalized phase-matching conditions for high harmonics: The role of field-gradient forces, Phys. Rev. A 55, 3204 (1997)

[26] H. Merdji, M. Kovačev, W. Boutu, P. Salières, F. Vernay, and B. Carré, Macroscopic control of high-order harmonics quantum-path components for the generation of attosecond pulses, Phys. Rev. A 74, 043804 (2006)

[27] P. Vaity, J. Banerji, and R. P. Singh, Measuring the topological charge of an optical vortex by using a tilted convex lens, Phys. Lett. A, 377, 1154-1156, (2013)

[28] C. M. Heyl, P. Rudawski, F. Brizuela, S. N. Bengtsson, J. Mauritsson, and A. L'Huillier, Macroscopic Effects in Noncollinear High-Order Harmonic Generation, Phys. Rev. Lett. 112, 143902 (2014)

[29] S. Kazamias et al., High order harmonic generation optimization with an aperture laser beam, Eur. Phys. J. D 21, 353-359 (2002)

[30] W. Boutu, T. Auguste, J. P. Caumes, H. Merdji, and B. Carré, Scaling of the generation of high-order harmonics in large gas media with focal length, PRA 84, 053819 (2011)

[31] E. Constant, D. Garzella, P. Breger, E. Mével, Ch. Dorrer, C. Le Blanc, F. Salin, and P. Agostini, Optimizing High Harmonic Generation in Absorbing Gases: Model and Experiment, Phys. Rev. Lett. 82, 1668 (1999)

[32] A. Camper, T. Ruchon, D. Gauthier, O. Gobert, P. Salières, B. Carré, and T. Auguste, High-harmonic phase spectroscopy using a binary diffractive optical element, PRA 89, 043843 (2014)

[33] Avner Fleischer, Ofer Kfir, Tzvi Diskin, Pavel Sidorenko and Oren Cohen, Spin angular momentum and tunable polarization in high-harmonic generation, Nature Photonics 8, 543-549 (2014)

[34] G. Lambert, B. Vodungbo, J. Gautier, B. Mahieu, V. Malka, S. Sebban, P. Zeitoun, J. Luning, J. Perron, A. Andreev, S. Stremoukhov, F. Ardana-Lamas, A. Dax, C.P. Hauri, A. Sardinha and M. Fajardo, Towards enabling femtosecond helicity-dependent spectroscopy with high-harmonic sources, Nature Communications 6: 6167 (2015)